# A QoS aware Novel Probabilistic strategy for Dynamic Resource Allocation

G Arun Kumar[a], Snehanshu Saha[a], Aravind Sundaresan[a], Bidisha Goswami[a]

[a]Department of Computer Science & Engineering, PESIT-BSC,Bangalore,India

**Abstract**

The paper proposes a two player game based strategy for resource allocation in service computing domain such as cloud, grid etc. The players are modelled as demand/workflows for the resource and represent multiple types of qualitative and quantitative factors. The proposed strategy will classify them in two classes. The proposed system would forecast outcome using a priori information available and measure/estimate existing parameters such as utilization and delay in an optimal load-balanced paradigm.

## 1. Introduction

Resource allocation with an optimally low delay and balanced loading is one of the major challenges in cloud service provisioning. Several research groups have worked in the domain of load balancing and traffic management. A service computing environment like a cloud or grid typically consists of more than one node or computation center. Various global and local distribution schemes are proposed by several researchers worldwide. The author [4] referred this problem as 'multi facility resource allocation problem' and attempted to find a cost function to address this. Once the server is selected the server is probed for accessing the service. Prior to avail such services, a typical business agreement takes place between these parties termed as Service Level Agreement. Service Level Agreement (SLA) is the only term between customer and service provider that governs the conditions to be fulfilled at the time of service delivery. A cloud service provider, offering the services from different servers, creates a schedule for the bundle of service as requested by the user with the terms laid down in SLA. Every cloud provider promises enormous elasticity of resource provisioning. Addressing the resource delivery within the timeline as per the metrics in SLA is a major challenge in any dynamic environment. The nature of the network is completely vulnerable and system strategy of servicing is heterogeneous. Researchers are working worldwide to make this more flexible, enhance the business factor and render the cloud more usability. The load balancing and optimization of existing resource is one of the most pursued topics in the area of distributed computing. Bio inspired model is one of the most discussed solutions. In [13] the authors proposed to optimize the large scale power distribution system by maximizing the utility function, inspired by population dynamics. The authors in [14] proposed a population dynamics model for data streaming over peer to peer network. Authors [15] have also proposed a predator-prey relation model to model the relationship between demand and supply of resources. There are very popular service schemes such as FCFS, round-robin, Largest job first etc. There exist several special scheduling algorithms [7] [8] [9] [10] [11] [12]. As per the authors of [6] all of these scheduling algorithms suffer from disk space management issue. Moreover, the state-space information creates small delay which affects overall service time span. This is the reason why randomization in resource availability might be a better approach. Game theory is one of the major approaches used for solving several optimization problems. The game theory as examined by the authors [1] in the environment of grid computing, where they experimented with an agent based cooperative, semi-cooperative and non-cooperative strategy. The author concluded that cooperation among the player is extremely important when multiple factors have to be considered. Authors [2] proposed a cloud bank model with bidding game strategy. Considering the situation of uncertainty of internet availability and the credit is assigned based on quality and constant online availability. Again, versatility of resources and appealing resource renting price including complexity is not viable to address every aspect of dynamic change here. Authors in [3] argued that cooperation among several cloud providers makes better profitability and optimization by proposing a stochastic linear programming game model. Authors in [5] proposed architecture for spectrum access in small cell network aiming to address heavy computational complexity.

**Our Approach**:

The game design proposed here is significantly different from the above. The model is conceived to forecast the probability of the expected game winner and casting a solution accordingly. Forecasting the outcome of the game depends on a-priori information for classification and the primitive that there exists a single fixed resource. The workflows/demands are the players for this game. The approach is to design the game based on previous game results. Since the resource is valuable, the proposed model will allocate the resource only to the player who wins the game. The model, designed as two sets of classes, contain the information about the components of the RI's (Resource Indicators) and exploit the history of winning the same game i.e. winning the resource to be allocated considering the present scenario of the RI's, strives to predict the outcome of the game.

## 2. Workflow Analysis

The scenario consists of several set of workflows demanding to occupy the same resource. This is conceptualized as a competition among two players bidding in a game. The winner would be granted the resource. This can be extended to a multiplayer game. The intention of the game is to forecast the winner with a-priori information available in the training set. The a-priori information is the vector indicating the outcome of the bidding results between the two players in previous "n" encounters. The game strategy computes the probability of winning or losing the game by extracting the features from the existing training set. The probability density function is assumed to be normally distributed.

The global solution to be attempted later is to approach the fixed resource as some property which several players lay their claims on. As a first iteration, the model is designed for two players. Let's assume the Players/workflows/jobs are C and D.

**Step 1**: The game is designed for grabbing a single resource. If the resource is allocated, then the player would be the winner. If not allocated, then the player would be loser.

**Step 2**: $RI -$ Resource Indicator is representative of an individual player. A random variable is allocated to represent $RI$. So $RI = 1$ if C wins and $RI = 0$ for D.

**Step 3**: We already have an a priori information.

Let's assume

$$X = [x_1, x_2, ..., x_n]$$

This implies C and D already played the game n times. X is the vector which contains the information about the outcome of the bidding between two players in the last n encounters. So, $x_3 = 1$ implies C wins and D loses the $3^{rd}$ bidding game. X is our feature vector

**Step 4**: We define a classifier function $Clf$ to calculate the value of $RI$ (to predict the outcome of a future bid for the same resource).

Therefore,

$$RI = Clf(X) = \begin{cases} 1 & if\ h_\theta(X) \geq 0.5 \\ 0 & if\ h_\theta(X) < 0.5 \end{cases}$$

where,

$[P(RI = 1 | X) = h\Box_\downarrow\theta (X) =)$ *the binary output variable denoting the winner of the future bid.*

Here the weight factors $\theta_1, \theta_2, ..., \theta_n$ are unknown factors, which is going to be estimated through a cost function.

To predict the future outcome a training set would be $[(X_1, RI_1),(X_2, RI_2),...,(X_m, RI_m)]$

## 3. Our Strategy

The problem consists of 2 players bidding to be allocated a single fixed resource. The a priori information is given in the form of the results of previous bidding encounters. In this case a priori information is a set of 15 results, each denoting a winner of the corresponding bid. The solution proposed in this model is: The a priori information is taken in the form of a feature vector which consists of the results of the last 15 encounters between the 2 players denoted by $X$. The classifier defined in the paper is based on a logistic regression classification model. The output of this classifier is $RI = 0 \; or \; 1$. If $RI = 1$, Player C wins and if $RI = 0$, Player D wins.

### 3.1. Solution Scheme

Regression Analysis is one of the most widely used techniques for analysing multifactor data. The reason behind using Regression Analysis is that it presents an equation that represents the relationship between a variable of interest (response variable) and a set of related predictor variables.

Given a response variable $y$ and predictor variables $X_1, X_2, X_3, ...., X_n$ we define the following relationship:

$$y = \theta_0 + \theta_1 x_1 + \theta_2 x_2 + \cdots + \theta_n x_n$$

This is called a multi-variable linear regression model which is parameterized by $\theta_1, \theta_2, ..., \theta_n$.

In linear regression the output given in the form of a response variable is continuous. The problem presented here, however requires the output in the form of a categorical variable. This presents the need to use Logistic Regression.

Logistic Regression is a statistical method for analysing a dataset in which there are one or more independent variables that determine an outcome. The outcome is measured with a dichotomous/binary variable.

Given a binary output variable RI, we model the conditional probability $P(RI = 1 | X = x)$ as a function of $x, p(x)$.

Formally, the logistic regression model is:

$$\log \frac{p(x)}{1 - p(x)} = \theta_0 + \theta.x$$

Logistic regression deals with this problem by using a logarithmic transformation on the outcome variable. It expresses the linear regression equation in logarithmic terms (called the logit).

Solving for $p$,

$$p(x; \theta) = \frac{1}{1 + e^{-(\theta_0 + \theta.x)}}$$

To ensure successful classification, we should predict $RI = 1$ when $p \geq 0.5$ and $RI = 0$ when $p < 0.5$. This implies guessing 1 when $\theta_0 + \theta.x$ is non-negative and 0 otherwise. Thus logistic regression gives us a linear classifier

## 3. Implementation

Given 2 players C and D bidding for a single resource, a classifier to predict the winner of the bid is built using logistic regression. The predictor variables $X = [x_1, x_2, \ldots, x_n]$ where $x_i$ is the outcome of the $i^{th}$ previous encounter between C and D i.e. 1 if C had won and 0 if D had won. $RI \in \{0,1\}$ is the binary output variable denoting the winner of the future bid.

Our classifier is formally defined using the following rule,

$$P(RI = 1|X) = h_\theta(X)$$

where,

$h_1\theta(X) =)$ *the binary output variable denoting the winner of the future bid.d won.e values.to*

(1)

Decide $RI = 1$ if $h_\theta(X) \geq 0.5$ else decide $RI = 0$.

Eq. 1 is parameterized by,

$$[\![\theta = [\theta]\!]_1, \theta_2, \ldots, \theta_n]$$

### 3.1. Parameter Estimation

To estimate $\theta$ we define the cost function,

$$J(\theta) = \begin{cases} -\log(h_\theta(X)) & if\ RI = 1 \\ -\log(1 - h_\theta(X)) & if\ RI = 0 \end{cases}$$

(2)

Given $m$ samples of training data,

$$[(X_1, RI_1), (X_2, RI_2), \ldots, (X_m, RI_m)]$$

Eq. 2 can be modified as,

$$J(\theta) = -\frac{1}{m}\left[\sum_{i=1}^{m} RI_i.\log(h_\theta(X_i)) + (1 - RI_i).\log(1 - h_\theta(X_i))\right]$$

Optimal $\theta$ is determined by,

$$\theta = \arg\min_\theta J(\theta)$$

The optimal $\theta$ obtained completely defines the classifier.

*3.2. Prediction*

Given an unlabelled sample $X$ , we assign the class label $RI = 0\ or\ 1$ using the classifier function $h_\theta(X)$.

*3.3. Algorithm*

The above mentioned methodology is implemented in MATLAB as follows:

1) Training:

**Input:** *X*, *RI* where

~~X~~          matrix of order *m* x *n*. *m* denotes the number of training samples. *n* is the number of features where each feature corresponds to a result in the previous 15 encounters between C and D.
Each row of *X* consists of a string of 0's and 1's where 1 denotes C is the winner and 0 denotes D is the winner.

~~RI~~          class label = 0 or 1.

**Output:** Optimal theta i.e. estimated parameters
- Initialize all the parameters to zero.

```
initial_theta = zeros (n + 1, 1)
```

- Define Sigmoid function

```
g = 1 / (1 + exp (-z))
```

- Define the Cost function to estimate theta

```
h_theta = sigmoid (X*theta)
J = (1/m)* sum ((-RI * log (h_theta))
 - ((1-RI) * log (1 - h_theta)))
```

- Compute the gradient

```
grad = (1 / m) * (h_theta - RI)' *X
```

- Use a general optimization technique to obtain optimal theta.
  In this case we have used **fminunc** function in MATLAB that uses a Trust - Region algorithm.

2) Testing:

**Input:** *X*

~~X~~          matrix of order m x n. m denotes the number of testing samples. n is the number of features where each feature corresponds to a result in the previous 15 encounters between C and D.

**Output:** *RI*
*RI* ~~—~~ class label for the test case.
If $RI = 1$ , Player C wins
Else if $RI = 0$ , Player D wins

- Define the Predictive function to determine the class label

```
RI = sigmoid (X * theta) >= 0.5;
```

## 4. Results and Discussion

We used the algorithm mentioned above to train the classifier with 1000 samples of the form:

$$[(X_1, RI_1), (X_2, RI_2), ..., (X_{1000}, RI_{1000})]$$

After training the classifier, the problem was simulated with n=1000, 5000 and 10000 where n is the number of job requests. The output of the classifier is a vector of labels, where each label denotes the winner of a future bidding encounter.

Comparing the output of the classifier with the actual class labels of the job requests we obtain the accuracy of the classifier. This denotes how reliable our classifier is in predicting the outcome of future job requests. The following accuracies were achieved after the simulation of the problem:

- For $n = 1000, Accuracy = 91.2\%$
- For $n = 5000, Accuracy = 91.08\%$
- For $n = 10000, Accuracy = 91.52\%$

Fig 1. Resource Utilization between Player C and Player D

The output of the classifier for a test case denotes the number of victories attained by each workflow. This information is then used to compute the resource utilization.

Resource utilization is computed to determine the amount of resource required by each workflow competing for it. The values of resource utilization tabulated below depict the percentage of the resource that must be allocated by the vendor for each of the workflows requesting the resource. As shown in Fig. 1, the overall utilization is 100%

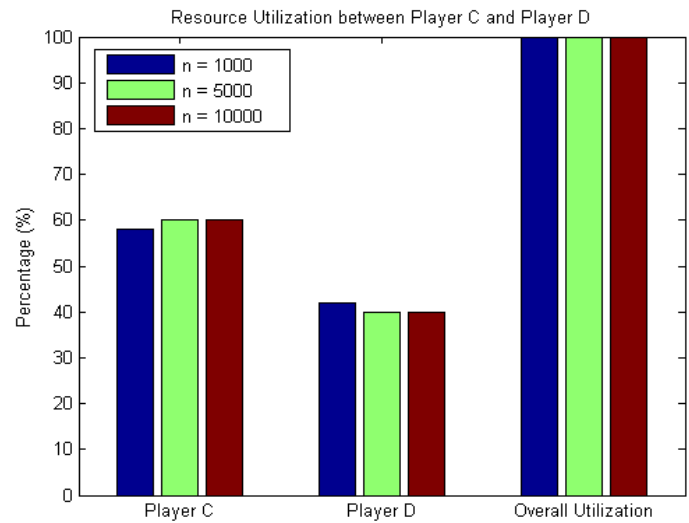

indicating negligible idle time.

The simulation results are tabulated as follows:

| n | Resource Utilization of | Resource Utilization of | Overall Utilization |
|---|---|---|---|

|       | C (%) | D (%) | (%)    |
|-------|-------|-------|--------|
| 1000  | 58.00 | 42.00 | 100.00 |
| 5000  | 60.10 | 39.90 | 100.00 |
| 10000 | 59.98 | 40.02 | 100.00 |

The resource utilization obtained for each workflow can be used by the service provider to formulate the Service Level Agreement (SLA) as it gives a rough estimate of the actual requirement of the resource by a customer.

There are existing scheduling algorithms like FCFS, Round-Robin and SJF that are used to manage resource allocation. An issue with these algorithms is that they require state-space information which creates a small delay which affects the overall time span. Our probabilistic approach eliminates the need for such prior information and henceforth the delay incurred is significantly reduced.

We represent the average delay per Job Request by a nonnegative random variable $D$ (Delay). A value $D_{\max}$ is defined which denotes a threshold on the random variable $D$.

We propose a **Delay estimation theorem**,

$$P(D \geq D_{\max}) \leq \frac{E[D]}{D_{\max}}; \quad D \geq 0, D_{max} > 0$$

**Proof**: Define a probability space $A = \{d \in \Omega \mid D(d) \geq D_{\max}\}$ where $\Omega \equiv$ Probability space that contains $A$ .

The mean value of $D$ ,

$$E[D] = \sum_{d \in \Omega} P(d).D(d)$$

$$= \sum_{d \in A} P(d).D(d) + \sum_{d \in \bar{A}} P(d).D(d)$$

(3)

The second term $\sum_{d \in \bar{A}} P(d).D(d) \geq 0$ by definition, therefore Eq. 3 becomes

$$E[D] \geq \sum_{d \in A} P(d).D(d)$$

(4)

We know that $D(d) \geq D_{\max}$ , therefore Eq. 4 becomes,

$$E[D] \geq D_{\max} \sum_{d \in A} P(d)$$

$$= D_{\max}.P(D \geq D_{max})$$

(5)

Eq. 5 can be rewritten as,

$$P(D \geq D_{\max}) \leq \frac{E[D]}{D_{\max}}$$

***QED***

If $D_{max} \gg D$ , then $P(D \geq D_{max})$ will be bounded by a small number. Hence the system will function within an acceptable delay. In fact this bound may be further tightened w.r.t accumulation of considerable historical data.

In the simulation the time required at each phase is listed as follows:

**Phase 1:** For Training, time consumed is 2.404 $s$

**Phase 2:** Predicting the outcomes of future job requests.

Let $D_{max} = 1 \, ms$ for the testing phase. The values of delay obtained for different test cases of the simulation are tabulated below:

| $n$ | Total Time required ($\mu s$ ) | Average Time per Request ($\mu s$ ) |
|---|---|---|
| 10 | 25.81 | 2.5810 |
| 50 | 26.13 | 0.5226 |
| 100 | 26.64 | 0.2664 |
| 500 | 119.54 | 0.2391 |
| 1000 | 132.91 | 0.1329 |
| 5000 | 176.97 | 0.0354 |
| 10000 | 230.48 | 0.0230 |

The following figure shows the relationship between the overall time delays for the various test cases.

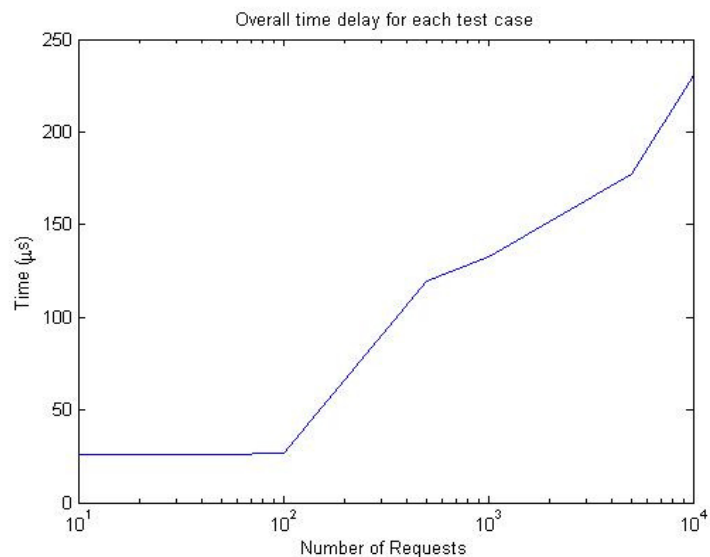

Fig 2. Overall Time Delay for each Test Case

Overall delay is increasing in piecewise linear fashion making the problem tractable and the delay of finite bounded variation.

Mean average time per job request,

$$E[D] = \frac{(2.5810 + 0.5226 + 0.2664 + 0.2391 + 0.1329 + 0.0354 + 0.0230)}{7}$$
$$= 0.5429 \, \mu s$$

Therefore, Eq. 5 becomes

$$P(D \geq 1 * 10^8) \leq \frac{0.5429}{1 * 10^8} = 0.5429 * 10^{-8}$$

The following figure shows the relationship between the average time delays for a Job Request for the various test cases.

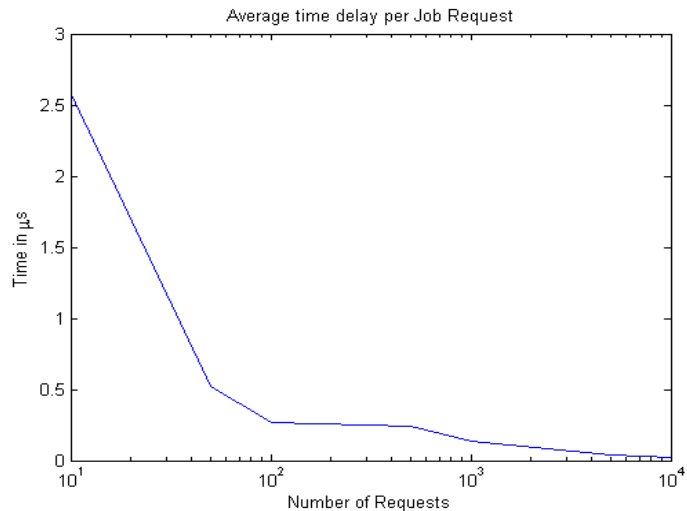

Fig 3. Average Time Delay per Job Request

The average delay estimation theorem concludes that the probability of average delay being greater than a threshold value is minimal implying that the average delay is finitely bounded above. The simulation validates the conclusion of the theorem by showing that average delay is low and stays in that range as the number of job requests increase.

## 5. Future Work

The proposed model which is initially designed for two players can be extended as a multiplayer game. In such a case the two class problem is elevated to a multi-class problem.
This work has already been extended to a multi-flow scenario with the Regression Logic edited and scaled up.
Accuracy is reported to be in the range of 75-80%.

Another possible dimension that can be added to the problem is the heterogeneity of the job requests. In this case multiple random variables $RI_1, RI_2, \dots, RI_k$ can be defined to represent $k$ different kinds of job requests.